\documentclass[12pt]{article}
\usepackage{helvet,times,mathptm}
\usepackage[square,sort,comma,numbers]{natbib}
\usepackage{graphicx}
\usepackage{multirow}
\usepackage{color}

\begin{document}
\begin{center}
{\noindent {\bf Studies on a frustrated Heisenberg spin chain with alternating
ferromagnetic and antiferromagnetic exchanges}}
\end{center}

\begin{center}
Shaon Sahoo,$^{a,}$\footnote[1]{shaon@sscu.iisc.ernet.in}
V. M. L. Durga Prasad
Goli,$^{a,}$\footnote[2]{gvmldurgaprasad@sscu.iisc.ernet.in}
Diptiman Sen$^{b,}$\footnote[3]{diptiman@cts.iisc.ernet.in}
and S. Ramasesha,$^{a,}$\footnote[4]{ramasesh@sscu.iisc.ernet.in}
\end{center}

\begin{center}
{\small \it
$^a$Solid State $\&$ Structural Chemistry Unit, Indian Institute of Science,
Bangalore 560 012, India \\
$^b$Centre for High Energy Physics, Indian Institute of Science,
Bangalore 560 012, India \\}
\end{center}

\begin{abstract}
{\noindent 
We study Heisenberg spin-1/2 and spin-1 chains with alternating
ferromagnetic ($J_1^F$) and antiferromagnetic ($J_1^A$) nearest-neighbor
interactions and a ferromagnetic next-nearest-neighbor interaction ($J_2^F$).
In this model frustration is present due to the non-zero $J_2^F$. The
model with site spin $s$ behaves like a Haldane spin chain with site spin
2$s$ in the limit of vanishing $J_2^F$ and large $J_1^F/J_1^A$.
We show that the exact ground state of the model can be found along a line
in the parameter space. For fixed $J_1^F$, the phase diagram in the space
of $J_1^A-J_2^F$ is determined using numerical techniques complemented by
analytical calculations. A number of quantities, including the structure
factor, energy gap, entanglement entropy and zero temperature
magnetization, are studied to understand the complete phase diagram. An
interesting and potentially important feature of this model is that 
it can exhibit a macroscopic magnetization jump in the presence of a
magnetic field; we study this using an effective Hamiltonian.}
\end{abstract}


\section{Introduction}
In nature matter comes in a plethora of quantum phases, each having its own
exotic properties. Normal metals, insulators, superfluids, superconductors, 
and different types of magnets are some of the many manifestations of quantum 
matter. Quantum spin systems with frustrating interactions often exhibit very 
rich phase diagrams at zero temperature~\cite{review1,review2,review3}.
Depending on the strength of the frustration and other parameters, such systems
can have magnetic or non-magnetic phases with or without a gap. Here we study 
one such interesting and important frustrated spin system in one dimension.

Our model is a one-dimensional spin model with alternating ferromagnetic
($J_1^F$) and antiferromagnetic ($J_1^A$) nearest-neighbor (NN) exchange
interactions. A next-nearest-neighbor (NNN) ferromagnetic exchange interaction
($J_2^F$) is also considered which induces frustration in the model. This 
model is important because it maps to the Haldane spin chain~\cite{haldane} 
in some limits. Besides its theoretical interest, the experimental 
realizations of this alternating spin model gave us additional impetus to 
study this model carefully. Some of the systems reported so far in this regard
are $[Cu(TIM)]CuCl_4$ \cite{hagi97}, $CuNb_2O_6$ \cite{koda99}, 
$(CH_3)_2CHNH_3CuCl_3$ \cite{mana97} and $(CH3)_2NH_2CuCl_3$ \cite{stone07}. 
Quite interestingly, we find in our study that the phase diagram of this model
with site spin 1/2 is very different from that of the model with site spin 1, 
even though both of them map to a integer spin Haldane chain in some limits.

Let us briefly mention here the work already done on this model. The spin-1/2
alternating model without NNN interactions ($J_2^F$ = 0) was studied as a
function of the strength (and sign) of one exchange interaction while the
other alternating interaction (antiferromagnetic) was kept fixed; it was shown 
how different phases, namely, a gapped Haldane phase, a gapped phase of 
decoupled singlets, a gapless spin-liquid phase and a gapped dimerized phase, 
appear as one varies the exchange parameter~\cite{hida92}. The excitation 
spectrum of this model ($J_2^F$ = 0) has been investigated by 
Hida~\cite{hida94}.
The exact ground state of the model along the ferromagnetic and non-magnetic
transition line has been studied by Dmitriev {\it et al}~\cite{dmitr97}. In
another work, Hida {\it et al} studied the model for an open chain by numerical
techniques and found different magnitudes of the edge spin in a region between
the ferromagnetic phase and the Haldane phase (where the edge spin is equal 
to 1/2)~\cite{hida12}. Nakamura studied this model with randomness in the 
exchange constants and $J_2^F = 0$~\cite{naka05}. The model with anisotropic
exchange interactions and $J_2^F = 0$ has been studied by Ren and 
Zhu~\cite{ren08}. This model has also been studied with on-site 
anisotropy~\cite{hida92a,oka96}. Kohmoto and co-workers have studied a 
$Z_2\times Z_2$ hidden symmetry of the model (without NNN interactions but 
with exchange anisotropy) and shown in particular that the symmetry is fully 
broken in the Haldane gapped phase~\cite{kohm92,yama93}.

It may be mentioned here that all these studies were done for spin-1/2 systems;
we have not found any detailed study of the frustrated alternating model
($J_2^F \ne 0$) with site spin 1. Here we study the model both for site
spin 1/2 and 1. Our results for the spin-1/2 system will help to verify many
of the previous results besides throwing new light on the underlying physics
of the model. Our study on spin-1 system reveals interesting new physics.
The phase diagram for the spin-1 model turns out to be quite different from
that of the spin-1/2 model. Different quantities, like the ground state spin,
energy gap, structure factor, entanglement entropy and Zeeman plots (zero
temperature energy level spectrum as a function of an applied magnetic field),
are studied in this work to understand the properties of this frustrated spin
model. We will also present an exact eigenstate of the model (for
any site spin $s$) for a particular value of $J_2^F/J_1^F$, and then prove
that for spin-1/2 and spin-1 systems this eigenstate is a ground state above
a critical value of $J_1^A$. We then conjecture that the eigenstate will be
a ground state of the model with any site spin when $J_1^A$ is larger than
some critical value which depends on $s$.

There exist two other types of widely studied frustrated spin 
chains. In one type of systems, both NN and 
NNN exchange interactions are antiferromagnetic in nature 
\cite{majumdar70, kumar02,chitra95,pati96,goli13,gerhardt98,kolezhuk12}. 
In the second type of frustrated systems, all the NN exchange 
interactions are ferromagnetic in nature while the NNN 
exchange interactions are antiferromagnetic in nature 
\cite{kolezhuk12,kolezhuk05,arlego11}. The frustrated system we study in this 
paper are not exactly equivalent to any of the above two frustrated 
systems as they cannot be mapped to each other in any limit.

Our paper is organized in the following way. In section 2, we discuss the
relevant Hamiltonian for the model. In section 3, we present and analyze 
the exact ground state of the model. In section 4, we present a classical 
analysis of the ground state phases of the model. In section 5, the quantum
phases of the model are studied in detail. This is done by extensive numerical
analysis complemented by some analytical calculations. In section 6, we 
study the zero temperature behavior of the system in an external magnetic 
field. We conclude our paper in section 7.

\section{The spin model}
\begin{figure}[]
\begin{center} \includegraphics[width=10.0cm]{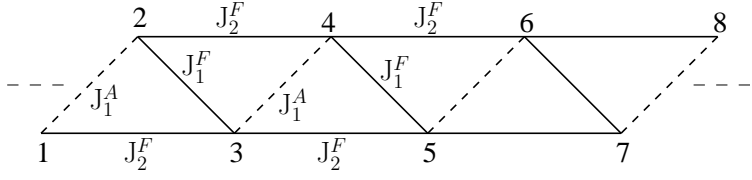}
\caption{A schematic diagram depicting the exchange interactions in the model
described by the Hamiltonian in Eq.~\ref{fafham}.} \label{systopo}
\end{center} \end{figure}

Our spin model is described by the following Hamiltonian:
\begin{eqnarray} H ~&=&~ J_1^A\sum_{k=1}^{N/2} \vec{\mathbf{s}}_{2k-1}
\cdot \vec{\mathbf{s}}_{2k} ~-~ J_1^F\sum_{k=1}^{N/2} \vec{\mathbf{s}}_{2k}
\cdot \vec{\mathbf{s}}_{2k+1}\nonumber \\
& & -~ J_2^F \sum_{l=1}^N \vec{\mathbf{s}}_{l}\cdot 
\vec{\mathbf{s}}_{l+2}. \label{fafham} \end{eqnarray}
with $J_1^A,~J_1^F,~J_2^F > 0$. Here $\vec{\mathbf{s}}_i$s are site spin
operators with spin value $s$,
$J_1^F$ ($J_1^A$) is the NN ferromagnetic (antiferromagnetic)
exchange constant and $J_2^F$ is the ferromagnetic NNN exchange constant.
The total number of spins, $N$, will always be taken to be even in our work.
A schematic diagram of this model is given in Fig. \ref{systopo}.

We will use periodic boundary conditions and will normalize
$J_1^F$ to 1 throughout this work. We will vary the value of $J_2^F$ from
0 to 1 to study the effect of frustration and vary the value of $J_1^A$ 
from 0.5 to 1.5 to study the effect of dimerization. We note that this is an 
interesting general model; with site spin $s$, it reduces to a Haldane chain 
with site spin 2$s$ in the limit of vanishing $J_2^F$ and large $J_1^F/J_1^A$. 

We mention here that all our numerical calculations are based on the exact 
diagonalization technique using Davidson's algorithm~\cite{davidson75}. The 
lowest energy state is calculated in each spin parity 
sector in the $M_S$=0 subspace to obtain the ground state as well as the spin 
gap. Further, we compute the expectation value of total $S^2$ in these states
by computing the expectation values of the two-point correlation functions 
${\vec S}_i \cdot {\vec S}_j$ between all pairs of sites $i$, $j$. This 
allows us to identify the total spin of these states.

\section{Exact ground state}
To the best of our knowledge there is only one kind of exact ground state
known so far for this frustrated alternating model. Later we will see that 
the entire phase diagram is divided into magnetic and non-magnetic phases; 
for a spin-1/2 system, Dmitriev {\it et al}~\cite{dmitr97} found the exact 
ground state along the transition line between
the phases. We will now show that for any $s$, an NN valence
bond singlet state is an eigenstate of the model when $J_2^F/J_1^F = 1/2$.

Let $[i,j]$ be the singlet state between spins at sites $i$ and $j$. We then
have the following relations: $\vec{ \mathbf{s}}_i \cdot \vec{ \mathbf{s}}_j 
[i,j] = -s(s+1)[i,j]$ and $\vec{ \mathbf{s}}_k \cdot(\vec{ \mathbf{s}}_i +
\vec{ \mathbf{s}}_j)[i,j] = 0$, for all $k~ \ne i, j$. Now, when
$J_2^F/J_1^F = 1/2$, we can rewrite the Hamiltonian in Eq.~\ref{fafham} in
the following form:
\begin{eqnarray} \mathbf{\tilde{H}}~&=&~ j_1^A \sum_{k=1}^{N/2}~ 
\vec{\mathbf{s}}_{2k-1}\cdot 
\vec{\mathbf{s}}_{2k}-\frac 12 \sum_{k=1}^{N/2}~\vec{\mathbf{s}}_{2k}
\cdot (\vec{\mathbf{s}}_{2k+1} + \vec{\mathbf{s}}_{2k+2}) \nonumber \\ 
&&~- \frac 12 \sum_{k=1}^{N/2}~\vec{\mathbf{s}}_{2k+1}\cdot 
(\vec{\mathbf{s}}_{2k} + \vec{\mathbf{s}}_{2k-1}), 
\label{splham} \end{eqnarray}
where $j_1^A$ is the new normalized NN antiferromagnetic exchange constant.
Using the above relations, it is easy to verify that the state $\psi = 
[1,2][3,4][5,6] \cdots [N-1,N]$ is an eigenstate of the Hamiltonian 
$\tilde{H}$, with
\begin{eqnarray} \tilde{H}~\psi &=& -\frac N2 s(s+1)j_1^A~\psi.
\label{egnst} \end{eqnarray}
We now prove that $\psi$ is a ground state of the system when $j_1^A$ is 
greater than a critical value which we will call $j_{1c}^A$. We use the 
following fact for this purpose. Suppose that the total Hamiltonian of a 
system is written as the sum of $M$ terms, i.e., $H = \sum_{i=1}^M H_i$. Using
the Rayleigh-Ritz variational principle, it can be proved that if a state is 
simultaneously a ground state of each of the $H_i$'s, then it will also
be a ground state of the total Hamiltonian. To apply this theorem in our
model, we decompose 
$\tilde{H}$ in Eq.~\ref{splham} as
$\tilde{H} = \sum_{k=1}^{N/2} (\tilde{H}_{2k-1} ~+~ \tilde{H}_{2k})$, where
$\tilde{H}_{2k-1} = \frac 12 j_A~ \vec{\mathbf{s}}_{2k-1}\cdot 
\vec{\mathbf{s}}_{2k} ~-~ \frac 12 \vec{\mathbf{s}}_{2k+1} \cdot 
(\vec{\mathbf{s}}_{2k} ~+~ \vec{\mathbf{s}}_{2k-1})$, and
$\tilde{H}_{2k} = \frac 12 j_A~\vec{\mathbf{s}}_{2k+1}\cdot 
\vec{\mathbf{s}}_{2k+2} - \frac12 \vec{\mathbf{s}}_{2k}\cdot 
(\vec{\mathbf{s}}_{2k+1} ~+~ \vec{\mathbf{s}}_{2k+2})$.
Here each of the parts corresponds to a block of three spins. All these block
Hamiltonians are essentially equivalent and have the same eigenvalues.

\begin{figure}[]
\begin{center} \includegraphics[width=9.5cm,height=7.0cm]{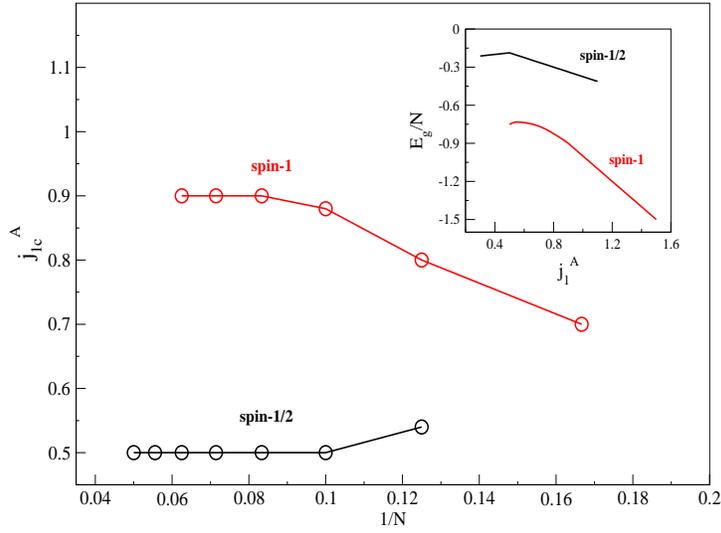}
\caption{Plot of $j_{1c}^A$ versus $1/N$ (here $J_2^F/J_1^F$ is 
fixed at 1/2). In the thermodynamic limit,
$j_{1c}^A$ saturates at $0.5$ for the spin-1/2 system and at $0.9$ for the
spin-1 system. In the inset, the ground state energy per site, $e_g = E_g/N$,
of the Hamiltonian $\tilde{H}$ is shown for a 20-site spin-1/2
chain and a 16-site spin-1 chain. The numerical value of $j_{1c}^A$ for
each chain length can be obtained by seeing where $e_g$ deviates from
its linear behavior or where the energy gap ($\Delta$) drops to minimum (see 
Fig. \ref{3lines}) as we reduce the value of $j_{1}^A$. The critical value
can also be estimated by noticing where the entanglement entropy jumps from 
zero to a non-zero value as we reduce the value of $j_{1}^A$ (see Figs. 
\ref{entcn_s0.5ring} and \ref{entcn_s1ring}).} \label{crtclval}
\end{center} \end{figure}

\begin{figure}[]
\begin{center} \includegraphics[width=12cm,height=9cm]{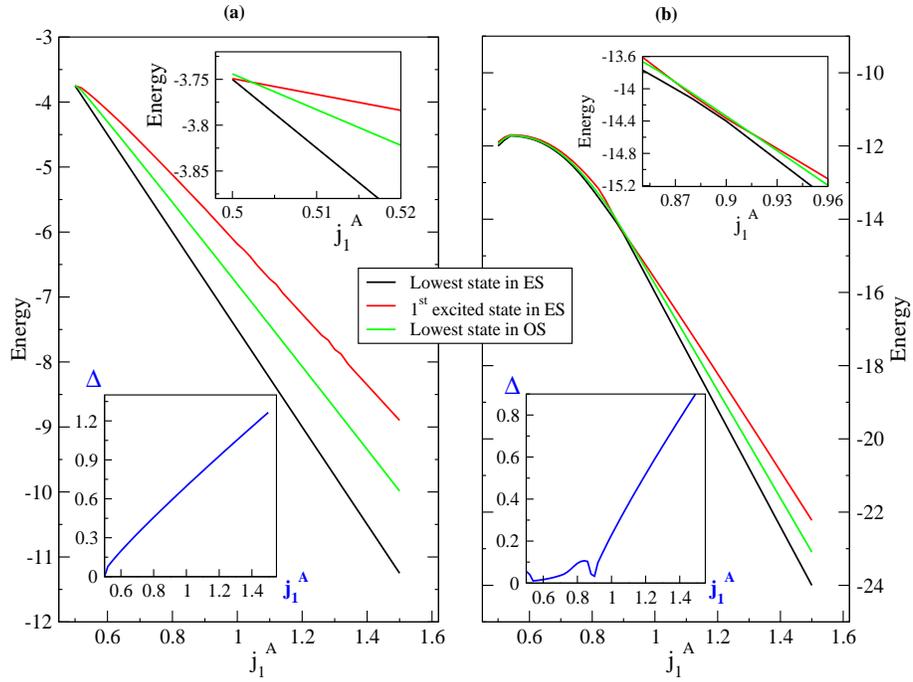}
\caption{Lowest two states from even spin parity space (ES) and lowest state
from odd spin parity space (OS) for a (a) 20 site spin-1/2 chain and (b)
16 site spin-1 chain (here $J_2^F/J_1^F$ is fixed at 1/2).
In the upper insets, the regions around the critical points are shown with 
more clarity. In the lower insets, the energy gaps ($\Delta$) are plotted 
against $j_1^A$.} \label{3lines} \end{center} \end{figure}

For a spin-1/2 system, each block Hamiltonian has the following three
eigenvalues (in the $S_z$ = 1/2 sector): $E_1 = -\frac{3j_1^A}{8}$, 
$E_2 = \frac{j_1^A}{8}-\frac14$ and $E_3 = \frac{j_1^A}{8} + \frac12$, with 
$E_1$ being the lowest one when $j_1^A\ge j_{1c}^A = 0.5$, below which $E_2$
becomes the lowest. Since the eigenvalue corresponding to the state
$\psi$ is equal to $NE_1$, we conclude that $\psi$ is a ground state of
$\tilde{H}$ (Eq.~\ref{splham}) when $j_1^A\ge j_{1c}^A$. The numerical value
we obtain for $j_{1c}^A$ in the spin-1/2 case is also $0.5$; below this the 
state $\psi$ is not the ground state (see Figs. \ref{crtclval} and 
\ref{3lines}(a)).

For a spin-1 system, each block Hamiltonian has the following seven 
eigenvalues (in the $S_z$ = 0 sector): $E_1 = -j_1^A$, $E_2 = -
\frac{j_1^A}{2}-\frac12$, $E_3 = -\frac{j_1^A}{2}+\frac12$, $E_4 = 
-\frac{j_1^A}{2}+1$, $E_5 = \frac{j_1^A}{2} - 1$, $E_6 = \frac{j_1^A}{2}+
\frac12$ and $E_7 = \frac{j_1^A}{2}+\frac32$, with $E_1$ being the lowest 
one when $j_1^A\ge j_{1c}^A = 1.0$, below which $E_2$
(and $E_5$ when $j_{1}^A < 0.5$) becomes the lowest. A similar argument
as given above establishes that $\psi$ is a ground state of
$\tilde{H}$ (Eq.~\ref{splham}) when $j_1^A\ge j_{1c}^A$.
The numerical value we obtain for $j_{1c}^A$ in the spin-1 case is $0.9$;
below this the state $\psi$ is not a ground state (see Figs. \ref{crtclval}
and \ref{3lines}(b)). Had we worked with bigger blocks (say, involving 
four spins), we surmise that our analytical value of $j_{1c}^A$ would have 
been closer to the numerical one.

The other NN singlet product state $[2,3][4,5][6,7] \cdots [N,1]$ is not an
eigenstate of the Hamiltonian $\tilde{H}$ and the ground state is
therefore unique. Following the results for $s$ = 1/2 and 1,
we conjecture that, for any $s$, the eigenstate $\psi$ will be a ground 
state of the Hamiltonian when $j_1^A$ is larger than some critical value. 
This critical value depends on $s$, and we expect it to increase with $s$.

Away from the critical point, the first excited state is expected to be a
triplet (Fig. \ref{3lines}) and can be expressed as a variational wave 
function in the space spanned by various low-energy excitations. Though 
finding the optimal wave function is non-trivial in general, the situation 
is much easier for large $j_1^A$. In this limit, the system behaves as a 
collection of $N/2$ isolated singlets, and as the energy cost to create a 
triplet by breaking an NN singlet is $j_1^A$, we expect that the excitation 
spectrum of the exact ground state will be gapped with the gap being 
$j_1^A$ (see Figs. \ref{gap_s0.5ring} and \ref{gap_s1ring}).

\section{Classical analysis of phase diagram}
In this section we analyze the classical phase diagram of the spin model
given in Eq.~\ref{fafham}; we will assume periodic boundary conditions.

 We begin by looking for classical ground states in which the 
angle between the spins at sites $n$ and $n+1$ is given by $\phi_1$ and 
$\phi_2$ for $n$ even and odd, respectively. Namely, $\vec{ \mathbf{s}}_{2n} 
\cdot \vec{ \mathbf{s}}_{2n+1} = s^2 \cos \phi_1$ and $\vec{ \mathbf{s}}_{2n-1}
\cdot \vec{ \mathbf{s}}_{2n} = s^2 \cos \phi_2$. We find that non-coplanar 
configurations have higher energy than coplanar ones for all values of the
parameters. Hence, we take all the spins to lie in the same plane; the angle 
between the spins at sites $n$ and $n+2$ must then be equal to $\phi_1 + 
\phi_2$ for all $n$, so that $\vec{ \mathbf{s}}_n \cdot \vec{ 
\mathbf{s}}_{n+2} = s^2 \cos (\phi_1 + \phi_2)$. In this configuration, the 
energy per site is given by (with $J_1^F$ normalized to 1)
\begin{eqnarray} e_0 ~&=&~ -\frac{s^2}{2} \cos \phi_1 ~+~ 
\frac{s^2 J_1^A}{2} \cos \phi_2 \nonumber \\
& & ~-~ s^2 J_2^F \cos (\phi_1 + \phi_2). \label{e0} \end{eqnarray}
Given some values of $J_1^A$ and $J_2^F$, we find the extrema of Eq.~\ref{e0} 
as a function of $\phi_{1,2}$. If an extremum occurs at angles denoted by
$\phi_{10}$ and $\phi_{20}$, we consider the matrix of second derivatives
around that point, $A_{ij} = (\partial^2 e_0/\partial \phi_i \partial 
\phi_j)_{\phi_{10},\phi_{20}}$. The extremum is a minimum if both the
eigenvalues of $A_{ij}$ are positive. A transition occurs from one
phase to another when one or both the eigenvalues of $A_{ij}$ crosses zero.

We then discover that there are four phases in the region with $J_1^A, J_2^F 
\ge 0$, in agreement with earlier work~\cite{dmitr97,hida12}.

\noindent (a) $\phi_1 = \phi_2 = 0$. This corresponds to a ferromagnetic phase
in which the spins are all parallel to each other. This phase lies in the 
region $J_1^A < 1$ and $J_2^F > \frac{J_1^A}{2(1 - J_1^A)}$.

\noindent (b) $\phi_1 = 0$ and $\phi_2 = \pi$. This corresponds to a
period-four configuration (double-period N\'{e}el phase with up-up-down-down
spin configuration). This phase lies in the region $J_2^F < \frac{J_1^A}
{2(1 + J_1^A)}$.

\noindent (c) $\phi_1 = \pi$ and $\phi_2 = \pi$. This corresponds to a
period-two configuration (N\'{e}el phase with up-down-up-down spin 
configuration). This phase lies in the region $J_1^A > 1$ and $J_2^F > 
\frac{J_1^A}{2(J_1^A - 1)}$.

\noindent (d) In the remaining regions, the classical ground state is given
by a spiral in which
\begin{eqnarray} \phi_1 &=& \cos^{-1} \left( \frac{\frac{1}{4} + (
\frac{J_2^F}{J_1^A})^2 - (J_2^F)^2}{\frac{J_2^F}{J_1^A}} \right), \nonumber \\
\phi_2 &=& \cos^{-1} \left( \frac{- \frac{1}{4} + (\frac{J_2^F}{J_1^A})^2 -
(J_2^F)^2}{J_2^F} \right), \label{phi12} \end{eqnarray}
with $0 < \phi_1 < \pi$ and $-\pi < \phi_2 < 0$.

As we move along the chain, the spins rotate by an average angle of
$(\phi_1 + \phi_2)/2$. This corresponds to a periodic configuration with
a wave length equal to
$4\pi/(\phi_1 + \phi_2)$. For a quantum spin chain with a large value of $s$,
i.e., in the semiclassical limit, this implies that the structure factor
$S(q)$, obtained by Fourier transforming the two-spin correlation function
${\vec s}_i \cdot {\vec s}_{i+n}$, will have a peak at a wave number given
by $q_{max} = (\phi_1 + \phi_2)/2$. In the four phases described above,
the peak will lie at $q_{max} = 0$, $\pi/2$, $\pi$ and $(\phi_1 + \phi_2)/2$
given by Eq.~\ref{phi12}, respectively.

\begin{figure}[]
\begin{center} \includegraphics[width=8.5cm,height=5.5cm]{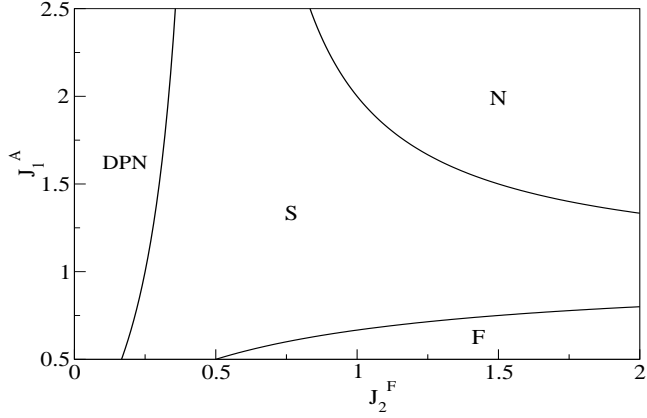}
\caption{Classical phase diagram of the spin model. The relative positions of
the N\'{e}el phase (N), spiral phase (S), ferromagnetic phase (F) and
double-period N\'{e}el phase (DPN) can be seen in the figure.} \label{clphdg}
\end{center} \end{figure}

The discussion of the phases above gives us the classical phase diagram of
the model. This can be seen from Fig. \ref{clphdg}. If we compare this 
classical phase diagram with the actual quantum phase diagram revealed later
in Figs. \ref{sf_s0.5ring} and \ref{sf_s1ring}, we see that, it correctly 
predicts the appearance of
the ferromagnetic phase at large $J_2^F/J_1^A$, though for somewhat different
parameter ranges. As one may expect, the phase diagram for higher spin
($s = 1$ shown in Fig. \ref{sf_s1ring}) resembles the classical phase diagram
more compared to that of the lower spin ($s = 1/2$ in Fig. \ref{sf_s0.5ring}).
In section 5 we study in detail the quantum phase diagram of the model.

It may be worth mentioning here that a spin wave analysis based
on one-magnon excitations predicts the same magnetic - non-magnetic transition
line, namely, $J_2^F = \frac{J_1^A}{2(1-J_1^A)}$ with $J_1^A < 1$, in agreement
with our classical analysis.

\section{Quantum phase diagram}
We have already seen the classical phase diagram of the model in 
Fig. \ref{clphdg}. 
In this section we will study the quantum phase diagram of this model. To 
understand the actual nature of the ground state, we first numerically 
calculate the ground state spin and the structure factor. Our results show a 
magnetic - non-magnetic transition, which is in agreement with the 
classical analysis. In the second part, we numerically calculate the energy 
gap and the entanglement entropy to understand the nature of excitation and 
the occurrences of quantum phase transition. As we will see, the study of 
the entanglement entropy also gives some new insights into the nature of the 
ground state.

\begin{figure}[]
\begin{center}
\includegraphics[width=10cm, height=6.0cm,
]{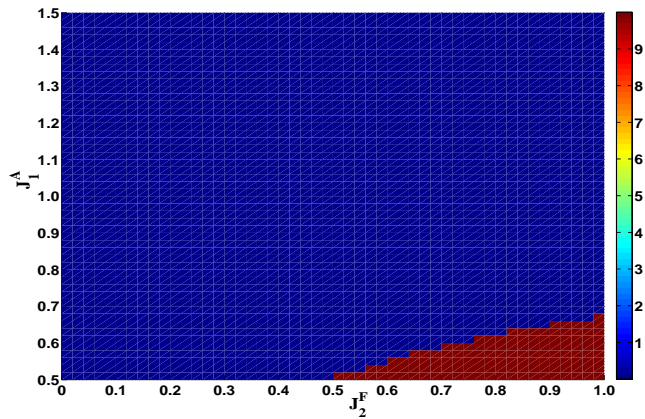}
\caption{Ground state spin as a function of $J_1^A$ and $J_2^F$ for a 20-site 
spin-1/2 chain. Calculations are carried out on a uniform grid of $101 
\times 101$.} \label{grspn_s0.5ring} \end{center} \end{figure}

\begin{figure}[]
\begin{center}
\includegraphics[width=10cm,height=6.0cm, 
]{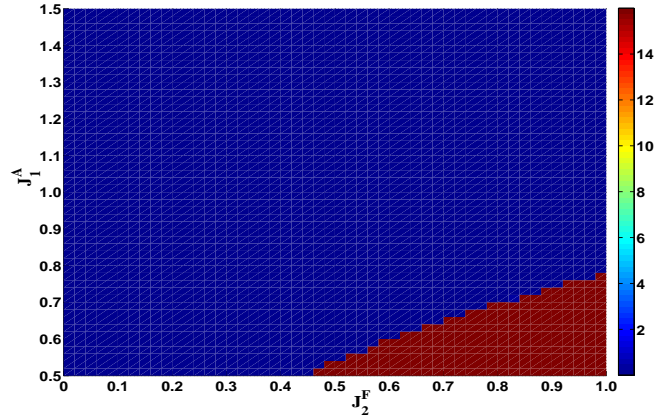}
\caption{Ground state spin as a function of $J_1^A$ and $J_2^F$ for a 16-site 
spin-1 chain. Calculations are carried out on a $101 \times 101$ uniform 
grid.} \label{grspn_s1ring} \end{center} \end{figure}

\paragraph{Ground state spin:}
The ground state spin has been calculated numerically for both spin-1/2 and 
spin-1 systems; the results can bee seen in Figs. 
\ref{grspn_s0.5ring} and \ref{grspn_s1ring} respectively. We see two totally 
different regions - one with total spin zero (non-magnetic or singlet ground 
state) and the other with maximum possible spin (ferromagnetic ground state). 
The existence of these two distinct phases can be understood in the following 
way. In the absence of frustration, when $J_1^F$ is large compared to $J_1^A$ 
in Eq.~\ref{fafham}, we expect the two spins connected by the ferromagnetic 
exchange to pair up in a symmetric combination and behave like a spin-$2s$ 
object. Then a weak antiferromagnetic interaction ($J_1^A$) would connect 
these coupled spins to form an effective antiferromagnetic chain with site 
spin $2s$. Hence, for the alternating spin-$s$ chain of length $N = 4n$ 
($n$ being an integer), the ground state will be a singlet.
This explains why our 20 site spin-1/2 system (Fig. \ref{grspn_s0.5ring}) 
and 16 site spin-1 system (Fig. \ref{grspn_s1ring}) show singlet ground 
states in the limit of vanishing $J_2^F$ and large $J_1^F/J_1^A$.

The exchange interaction $J_2^F$ in our spin model acts between two
NN coupled spins of the effective antiferromagnetic chain
occurring in the large $J_1^F/J_1^A$ limit. This implies that there is a
competition between two opposite interactions ($J_2^F$ and $J_1^A$) acting
between two neighboring coupled spins of the effective antiferromagnetic 
chain. As a result, as $J_2^F$ increases we expect to get a ground state 
with non-zero spin. This explains the existence of the ferromagnetic phase in 
Figs. \ref{grspn_s0.5ring} and \ref{grspn_s1ring}.

\begin{figure}[]
\begin{center}
\includegraphics[width=10cm,height=6cm, 
]{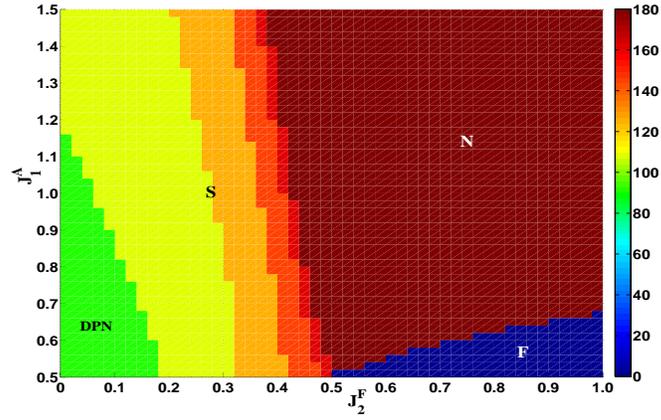}
\caption{Value of $q$ in degrees at which the structure factor is maximum as 
a function of $J_1^A$ and $J_2^F$ for a 20-site spin-1/2 chain. 
Calculations are carried out on a $51 \times 51$ uniform grid.} 
\label{sf_s0.5ring} \end{center} \end{figure}

\begin{figure}[]
\begin{center}
\includegraphics[width=10cm,height=6cm, 
]{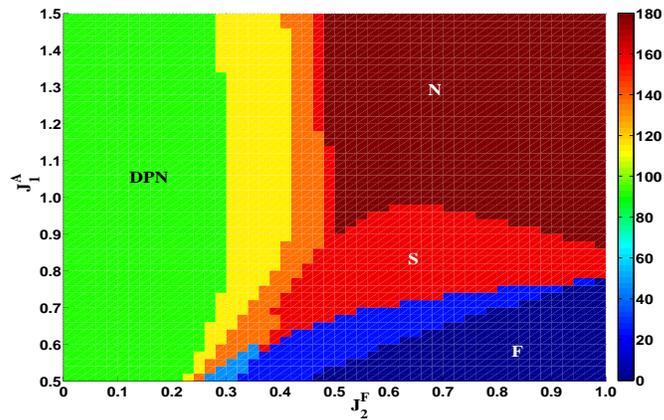}
\caption{Value of $q$ in degrees at which the structure factor is maximum as 
a function of $J_1^A$ and $J_2^F$ for a 16-site spin-1 chain. 
Calculations are carried out on a $51 \times 51$ uniform grid.} 
\label{sf_s1ring} \end{center} \end{figure}

\paragraph{Structure factor:}
To understand the exact nature of the `spin orientation' in different regions 
of the phase diagram, a numerical calculation is carried out to find the wave 
vector $q$ at which the modulus of the structure factor $S(q)$ 
is maximum. $S(q)$ is defined as the Fourier transform, 
$\sum_{n=0}^{N-1}{\rm e}^{iqn}\langle \vec{\mathbf{s}}_1 \cdot
\vec{\mathbf{s}}_{1+n}\rangle$, with $q$ going from $-\pi$ to $\pi$ in steps 
of $2\pi/N$. As argued in section 3, this value of $q$ gives us an idea 
about the relative orientation of the spins in space in the classical limit 
of the model. The results can be seen in Figs. \ref{sf_s0.5ring} and 
\ref{sf_s1ring} respectively for spin-1/2 and spin-1 systems. 
We see from these figures that there are two distinct quantum phases, a 
ferromagnetic phase (F) with $q_{max}=0$ and a non-magnetic phase with 
$q_{max}>0$. The non-magnetic phase, in turn, consists of three different 
regions, N\'{e}el (N) with $q_{max}=\pi$, double-period N\'{e}el (DPN) with 
$q_{max}=\pi$/2, and spiral (S) with $0<q_{max}<\pi$ ($q_{max}\ne\pi/2$). In 
the classical limit, the ferromagnetic and first two non-magnetic regions
correspond to the following spin configurations:
all spins are parallel (F), up-down-up-down (N) and up-up-down-down (DPN).
The system size dependence of the phase diagrams has also been
studied by carrying out calculations on a coarse grid of $10 \times 10$. For 
the spin-1/2 case, we have studied system sizes from 16 to 28; with increasing
system size we find that the DPN region extends further up along the $J_1^A$ 
axis than is shown in Fig. \ref{sf_s0.5ring} for a 20-site system. For the 
spin-1 case, we have studied system sizes from 8 to 16; we find that the 
spiral region lying between the N\'{e}el region (N) and the ferromagnetic 
phase (F) shrinks with increasing system size.

\begin{figure}[]
\begin{center}
\includegraphics
[width=16.0cm]{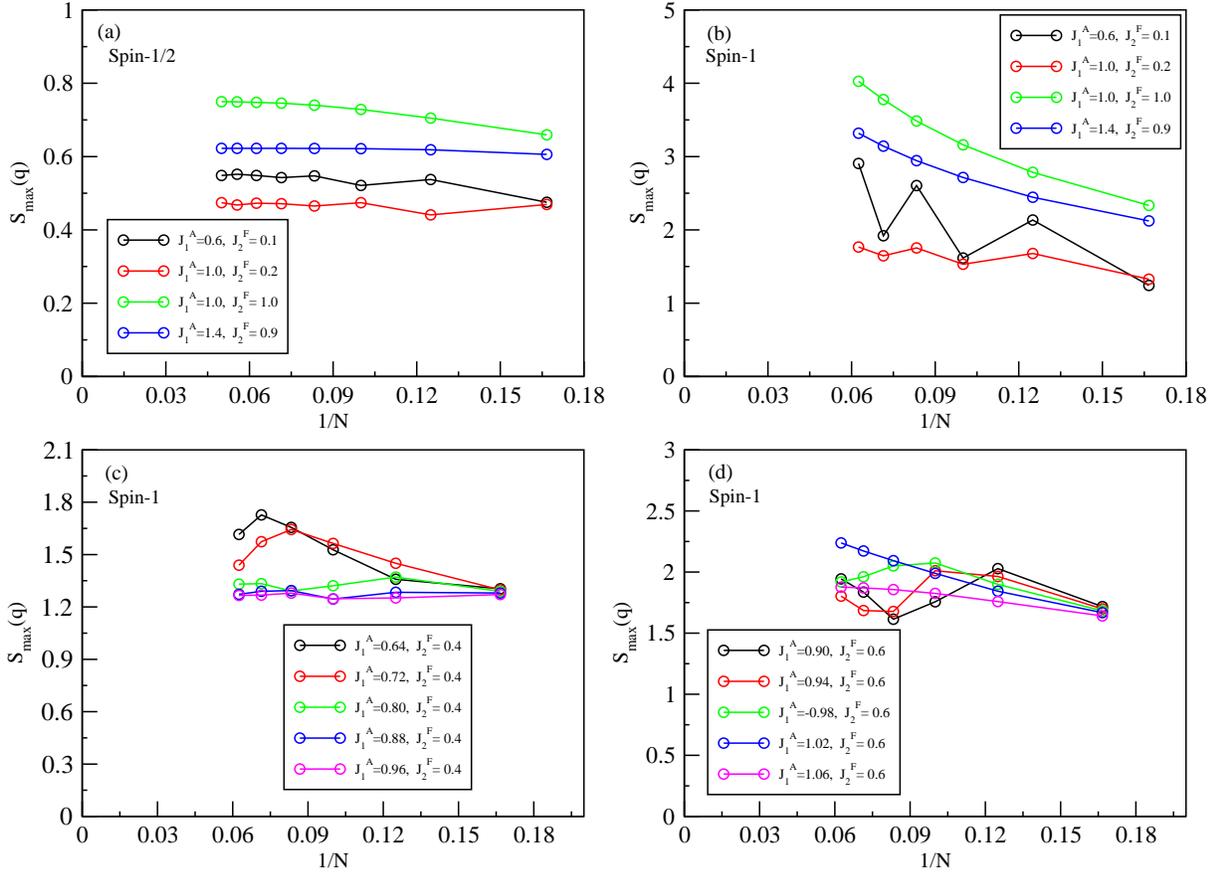}
\caption{ The maxima of $S(q)$ is plotted against $1/N$ for the 
different points in the phase diagram (see Figs. \ref{sf_s0.5ring} and 
\ref{sf_s1ring} respectively for s=1/2 and s=1 systems). In all the cases, 
$S_{max}(q)$ does not appear to have a very large value in the thermodynamic 
limit. The site spins of the model under consideration are shown in the 
figures.} \label{sqmax} \end{center} \end{figure}

We would like to emphasize that the spin-1/2 and spin-1 systems have no 
long-range order in any of the non-magnetic phases. In fact, the 
Mermin-Wagner theorem precludes the spontaneous breaking of a continuous
symmetry and hence long-range antiferromagnetic or spiral order in 
one-dimensional spin chains, even at zero temperature. Consequently,
in the limit $N \to \infty$, the peaks in the structure factor have finite 
heights and widths instead of being $\delta$-functions. (These features of the
structure factor have have been studied in detail in the spin-1/2 chain with 
NN and NNN antiferromagntic interactions in Ref. \cite{bursill} and generally 
for spin chains near valence bond solid points in Ref. \cite{nomura}).
In Fig. \ref{sqmax}, the maxima of $S(q)$ is plotted 
against $1/N$. Both for linear and quadratic fitting, the maxima of $S(q)$ 
does not seem to go to very large values in the thermodynamic limit. 
The notation that we have used for the different regions, N\'eel, 
double-period N\'eel and spiral, are based on the classical analysis of the 
phase diagram; they are convenient for distinguishing between the different 
regions, but they are not meant to imply that the quantum systems have 
long-range order or that they are truly different phases. The different 
non-magnetic regions are separated from each other by cross-over regions 
rather than phase transition lines (unlike the classical phase diagram). 
The only true quantum phase transition in the quantum systems appears to be
the one between the ferromagnetic phase and the non-magnetic phase, 
although for the spin-1 system we speculate an additional phase transition 
line within the non-magnetic phase (this will be discussed with 
Figs. \ref{entcn_s1ring}, \ref{gap_ja} and \ref{engap_ab} below).

We note that the extreme left part of the double-period N\'eel region, 
where $J_2^F$ is small, is similar in character to the Haldane phase of 
antiferromagnetic integer spin chains~\cite{haldane}. This follows from the
structure of Eq.~\ref{fafham}; the ferromagnetic term proportional to 
$J_1^F$ makes the pair of spins at sites $2k$ and $2k+1$ combine to form 
a spin-$2s$ object (i.e., an object with integer spin regardless of
whether $s$ is an integer or half-odd-integer), and each of these objects 
then have antiferromagnetic interactions with their neighboring objects 
due to the $J_1^A$ term.

\paragraph{Energy gap and entanglement entropy:}$ $ 
The entanglement entropy is now a well known tool to study and characterize 
the quantum many-body systems \cite{vedral08,amico08}. The behaviors of the 
entanglement entropy and energy gap of a spin system are interrelated and 
complementary, and one gives information about the other \cite{goli13}. 
The rate of change of entropy is known to be high wherever the energy gap is 
negligibly small in the phase diagram. In this subsection we will carry 
out a combined study of the energy gap, $\Delta$, and the 
entanglement entropy, $S_e$, to investigate the nature of the excitation 
spectrum, i.e., whether there are some gapless regions in the otherwise 
predominantly gapped non-magnetic phase. This study will reveal the
occurrence of possible quantum phase transitions in the phase diagram. 

The actual value of entropy also gives some useful insight into 
the character of a state. If entropy is low, we expect that the ground state is 
predominantly Kekule in nature (where neighbors form singlet pairs). If the 
entropy is high, we expect the ground state to have considerable contributions 
from the basis states with distant neighbor singlets. All these features of 
the entropy in the context of many-body systems can be explained within the 
framework of a valence bond theory \cite{goli13,sahoo12}, 
which is beyond the scope of the present work.

Before we present the results, let us briefly mention here how 
one calculates the entanglement entropy. A pure state of a bipartite system 
(divided into left and right blocks) can be written as
$\displaystyle|\psi\rangle = \sum_{ij} C_{ij} |\phi_i\rangle^l |\phi_j
\rangle^r$, where $ |\phi_i\rangle^l$ and $|\phi_j\rangle^r$ are the basis
states of the left and right blocks respectively. The reduced density matrix
(RDM) of the left block, $\rho_l=Tr_r(|\psi\rangle \langle\psi|)$, is
calculated by tracing out the degrees of freedom of the right block.
The elements of the RDM $\rho_l$ are given by
\begin{eqnarray} \rho_{ij} = \sum_k C_{ik}C_{jk}^* \label{rho1}. 
\end{eqnarray}
The von Neumann entropy of a block is given by 
$ S_e=-Tr(\rho~{\rm log}_2~\rho )$ or
\begin{eqnarray} S_e = -\sum_i ~\lambda_i~{\rm log_2}~\lambda_i, \label{entrp2} 
\end{eqnarray}
where the $\lambda_i$'s are the eigenvalues of $ \rho$.

\begin{figure}[]
\begin{center}
\includegraphics[width=10cm,height=6cm, 
]{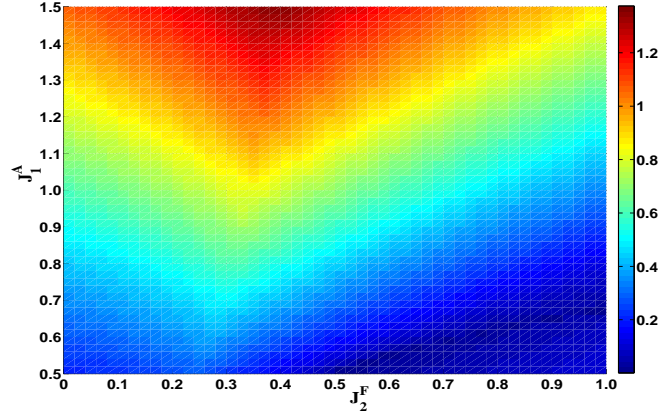}
\caption{Energy gap ($\Delta$) between the lowest two states as a function of 
$J_1^A$ and $J_2^F$ for a 20-site spin-1/2 chain. Calculations 
are carried out on a $101 \times 101$ uniform grid.} \label{gap_s0.5ring}
\end{center} \end{figure}

\begin{figure}[]
\begin{center}
\includegraphics[width=10cm,height=6cm,
]{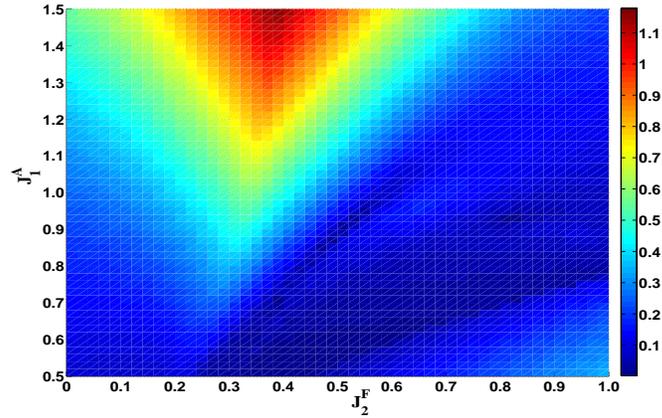}
\caption{Energy gap ($\Delta$) between the lowest two states as a function of
$J_1^A$ and $J_2^F$ for a 16-site spin-1 chain. Calculations are 
carried out on a $101 \times 101$ uniform grid.} \label{gap_s1ring}
\end{center} \end{figure}

\begin{figure}[]
\begin{center}
\includegraphics[width=10cm,height=6cm,
]{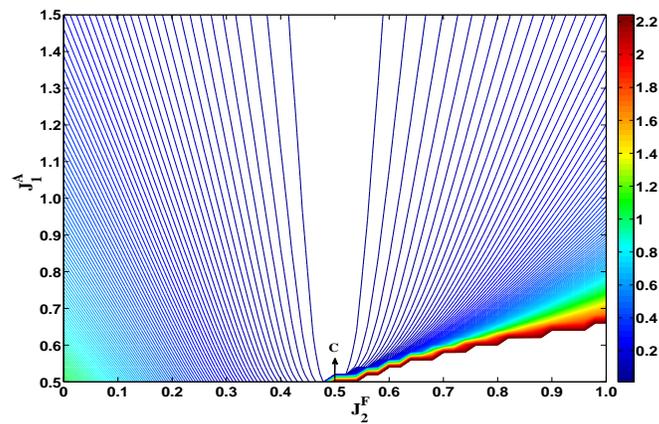}
\caption{Contour diagram of entanglement entropy ($S_e$) as a function of
$J_1^A$ and $J_2^F$ for a 20-site spin-1/2 chain. Calculations are carried out
on a $101 \times 101$ uniform grid.} \label{entcn_s0.5ring} \end{center} 
\end{figure}

As the first step of our study, we numerically calculate $\Delta$ (between the
lowest two states) as a function of of $J_2^F$ and $J_1^A$. The results can be
seen in Figs. \ref{gap_s0.5ring} and \ref{gap_s1ring} respectively for spin 1/2
and 1 systems. As expected, we see that the energy gap is very small in the 
ferromagnetic phase; the small non-zero values are due to finite size effects.
Interestingly, for large $J_1^A$ and intermediate range of $J_2^F$, one gets a
phase with large $\Delta$ and non-degenerate ground state. To understand this 
behavior and other parts of the phase diagram in the non-magnetic region, we 
analyze the bipartite entanglement entropy of the system.
To obtain the entropy we divide the full system into two equal parts in such
a way that the boundary of the partition cuts two ferromagnetic bonds. The
value of $S_e$ as well its rate of change with the exchange parameters can 
be seen from contour diagrams given in Figs. \ref{entcn_s0.5ring} and 
\ref{entcn_s1ring} respectively for spin 1/2 and 1 systems. 
 
\begin{figure}[]
\begin{center}
\includegraphics[width=10cm,height=6cm,
]{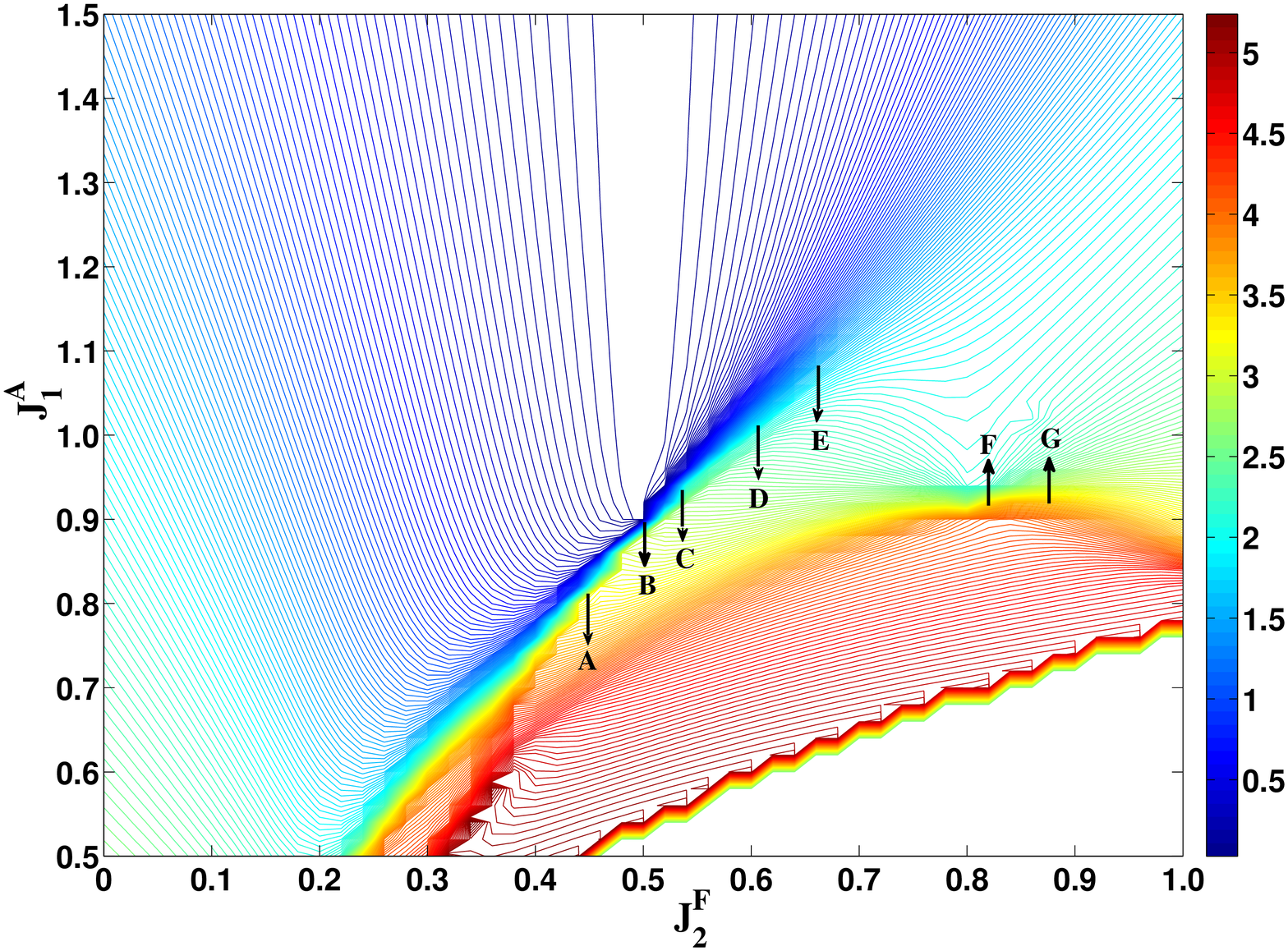}
\caption{Contour diagram of entanglement entropy ($S_e$) as a function of
$J_1^A$ and $J_2^F$ for a 16-site spin-1 chain. Calculations are carried out 
on a $101 \times 101$ uniform grid.} \label{entcn_s1ring} \end{center} 
\end{figure}

\begin{figure}[]
\begin{center} \includegraphics[width=12.6cm,height=9.0cm]{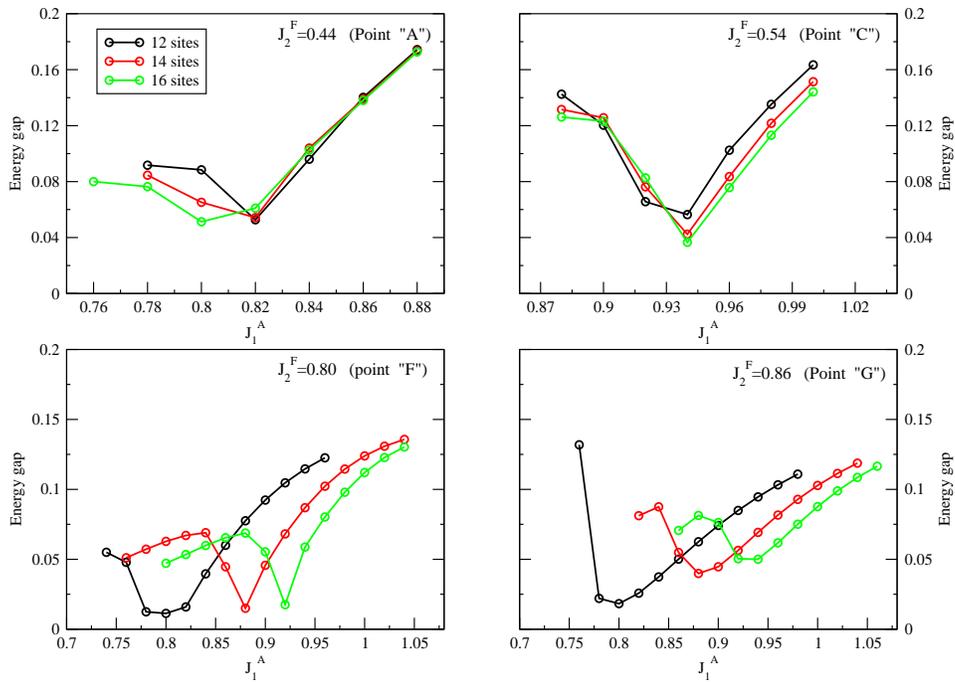}
\caption{Energy gap minimum at some representative points (marked in Fig. 
\ref{entcn_s1ring}) for various system sizes (system: spin-1 
chain)}. \label{gap_ja} \end{center} \end{figure}

The figures show that there exists a very low entropy flat region in the
above mentioned limit (large $J_1^A$ and intermediate range of $J_2^F$). 
The low entropy indicates that the ground state is a Kekule state with 
singlets between two neighboring sites coupled by $J_1^A$. This is supported 
by our exact result for $J_2^F= 1/2$, where the ground state is a product of
NN singlets for $J_1^A$ larger than some critical value. As the excitation 
requires the breaking of a singlet bond, this state is gapped. As discussed 
earlier, $\Delta$ increases linearly with $J_1^A$ in the large $J_1^A$ limit.
However this does not explain why the region of entropy minima is not
exactly at the place in the phase diagram where $\Delta$ is maximum.

From Fig. \ref{entcn_s1ring}, we can speculate whether there are any quantum
phase transitions (QPT) taking place. We know that the entropy susceptibility
(rate of change of entropy with respect to a given parameter)
is a good tool to detect QPTs. It can be seen from the figure that there are
regions in the non-magnetic phase where the entropy susceptibility has peaks, 
i.e., the density of contour lines is high. Following our previous study of 
the entropy contour diagram for spin systems~\cite{goli13}, we recognize here 
two different dense patterns of contour lines corresponding to two
different phases. The dense line passing through the points ``A", ``B", ``C",
``D" and ``E" is speculated to be gapless, which is supported by the finite
size analysis (Figs. \ref{gap_ja} and \ref{engap_ab}). The other dense line 
passing through the points ``F" and ``G" has a strange behavior. The energy 
gap goes through a minimum while crossing the line, but that minimum does not 
seem to come down to zero in the thermodynamic limit. In fact, our numerical 
results show (Fig. \ref{gap_ja}) that the energy gap minimum increases with 
the system size and appears to saturate in the thermodynamic limit.

\begin{figure}[]
\begin{center} \includegraphics[width=12.0cm,height=8.4cm]{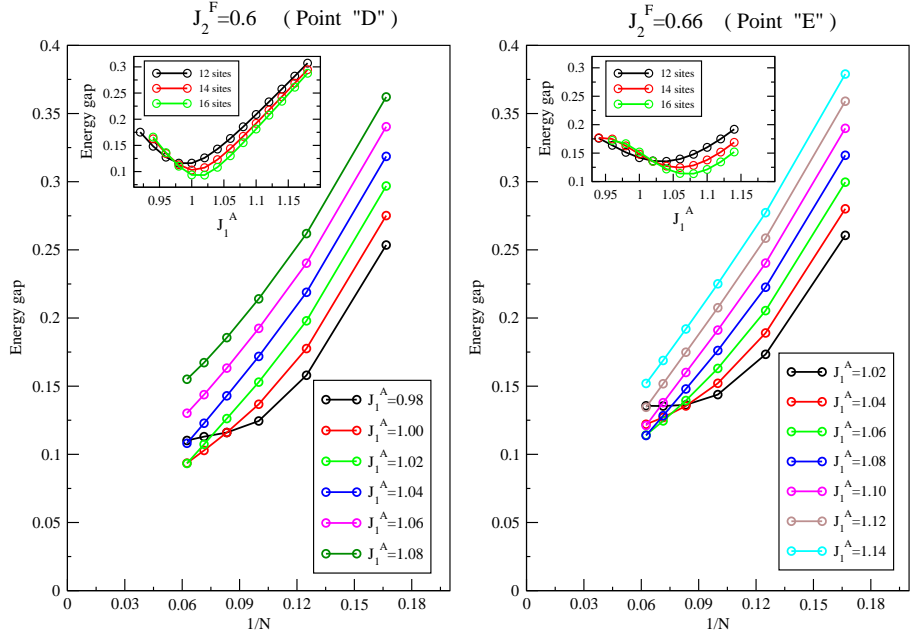}
\caption{Variation of the energy gap with the system size for different values
of $J_1^A$ for some fixed values of $J_2^F$
(system: spin-1 chain). For a given point, ``D" or ``E"
(marked in Fig. \ref{entcn_s1ring}), it seems that in the thermodynamic limit 
the energy gap may go to zero for a particular value of $J_1^A$. In the 
insets, it is clearly shown how the energy gap minima corresponding to a point
changes with system size. It can be seen that the gap minima decreases with 
increasing system size.} \label{engap_ab} \end{center} \end{figure}

For the spin-1/2 system, it is found that the entire non-magnetic region
is gapped (Fig. \ref{gap_s0.5ring}), except probably the region near the 
magnetic - non-magnetic transition line, where the entropy susceptibility is 
high (Fig. \ref{entcn_s0.5ring}).

\section{Macroscopic magnetization jump}
\begin{figure}[]
\begin{center} \includegraphics[width=12.0cm,height=6.3cm]{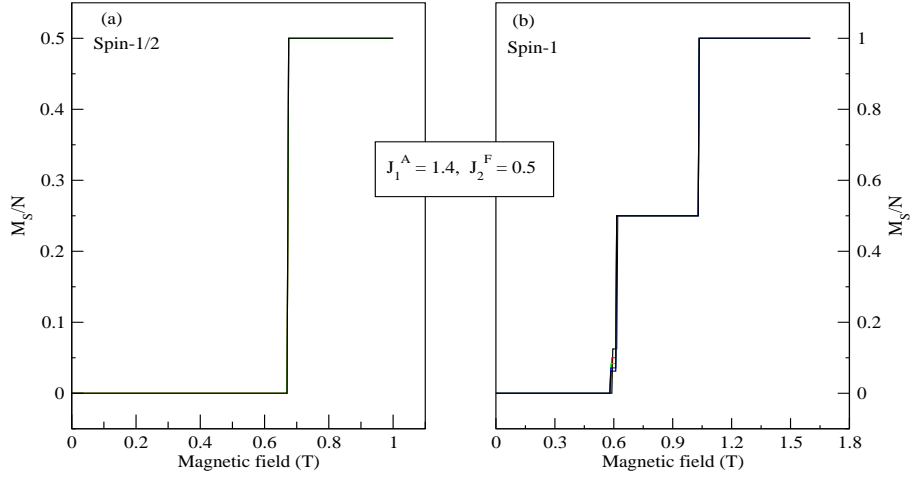}
\caption{ Zero temperature magnetization or Zeeman plot: (a) 
macroscopic jump in magnetization of a spin-1/2 system with different system 
sizes (N = 14 to 20) (b) macroscopic jump in magnetization of a spin-1 system 
with different system sizes (N = 8 to 16). 
Here the magnetization is normalized to the system sizes; the universal 
character of the jumps are evident from the plots.} 
\label{zee} \end{center} \end{figure}

\begin{figure}[]
\begin{center}
\includegraphics[width=10cm,height=6cm, 
]{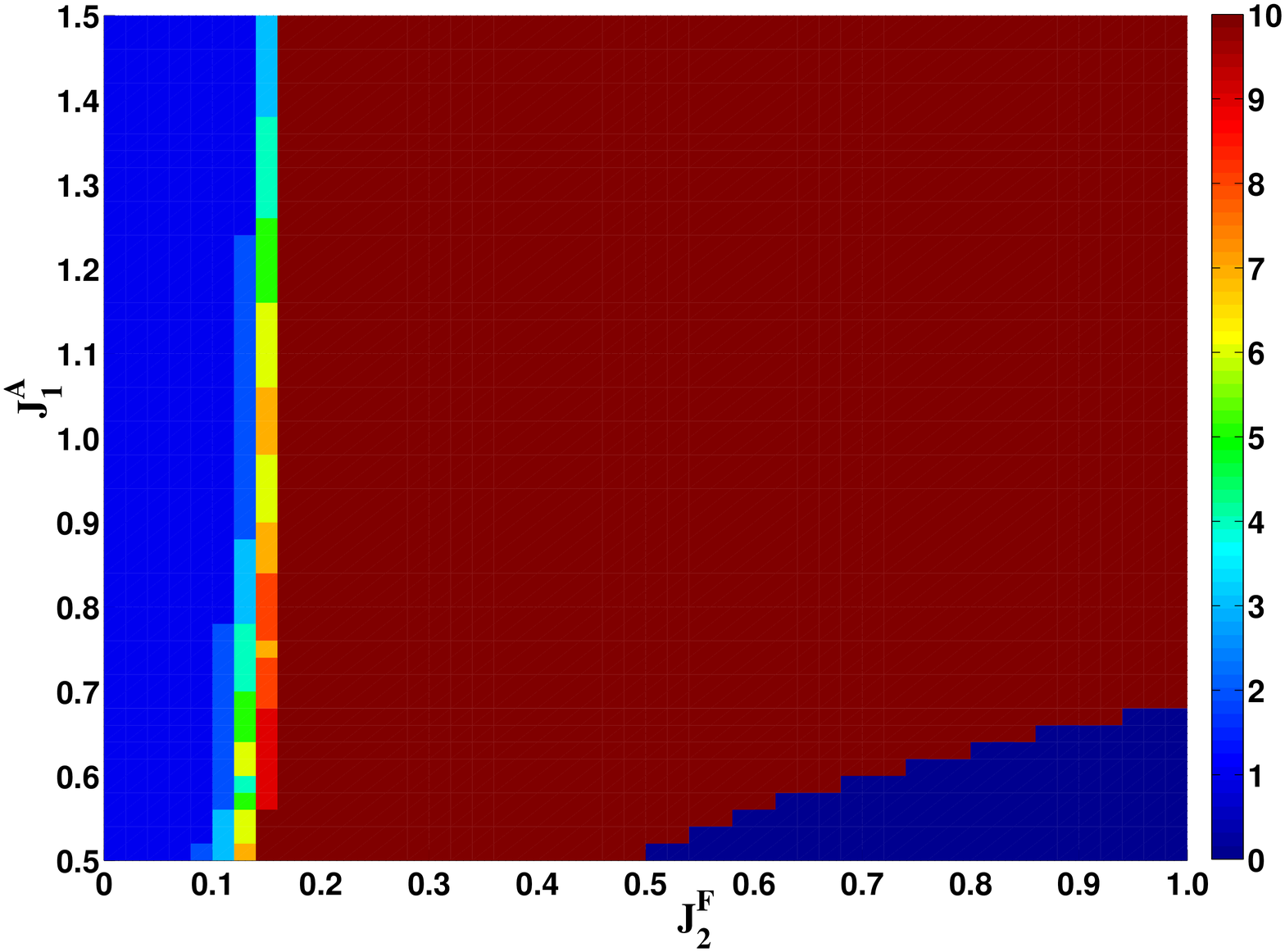}
\caption{Maximum jump in the magnetization as a function of $J_2^F$ and
$J_1^A$ for a 20-site spin-1/2 chain. Calculations are carried 
out on a $101 \times 101$ uniform grid.} \label{maxjmp_s0.5ring} 
\end{center}
\end{figure}

\begin{figure}[]
\begin{center}
\includegraphics[width=10cm,height=6cm, 
]{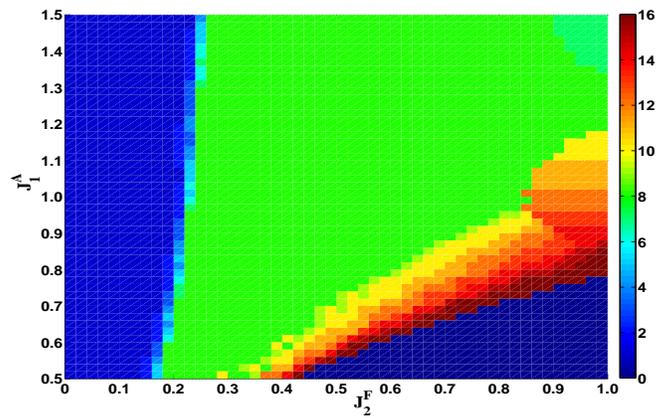}
\caption{Maximum jump in the magnetization as a function of $J_2^F$ and
$J_1^A$ for a 16-site spin-1 chain. Calculations are carried out on a $101 
\times 101$ uniform grid.} \label{maxjmp_s1ring} \end{center} \end{figure}

In this section we study the zero temperature magnetization of the system to 
understand how the lowest energy levels corresponding to different $S^z$ 
sectors are ordered with respect to each other. In the absence of a magnetic 
field, the magnetization is zero in the non-magnetic region. As one increases 
the magnetic field (applied along, say, the $z$-axis), the energy levels will
start to shift linearly with the field, as $-g\mu_B h S^z$, where $g$,
$\mu_B$ and $h$ are respectively the gyromagnetic ratio ($\approx$ 2), Bohr 
magneton and applied magnetic field. Since the rate of change in the energy 
levels depends on their $S^z$ values, we will get magnetization jumps as $h$ 
is increased depending on the energy gaps between the consecutive lowest 
energy levels and their respective $S^z$ values. Though the magnetization 
generally increases in steps of 1, in the
limit of large $J_1^A$ we see some big jumps (Fig. \ref{zee}). Since such 
large jumps in the magnetization are interesting and important, we have 
obtained the maximum jump corresponding to each point in the phase diagram 
for both spin 1/2 and 1 systems; the results can be seen in Figs. 
\ref{maxjmp_s0.5ring} and \ref{maxjmp_s1ring} respectively.

We now present a way of understanding the large jumps in the magnetization of 
the system as the magnetic field is varied. These jumps are macroscopic, 
namely, the magnetization $M= \sum_n s_n^z$ changes by a finite fraction of 
the total number of spins $N$; they have been studied earlier in the context 
of frustrated spin systems~\cite{schulenberg}. We will study this phenomenon 
using the idea of an effective Hamiltonian~\cite{totsuka,mila,kolezhuk,tandon}.

In the presence of a magnetic field applied along the $z$ direction, we
have to add a term equal to $-g\mu_B h \sum_n s_n^z$ to the Hamiltonian in
Eq.~\ref{fafham}. To obtain an effective Hamiltonian, we will assume that
the antiferromagnetic coupling $J_1^A$ is much larger than the other two
couplings given by $J_1^F = 1$ and $J_2^F$. We then write the Hamiltonian as
\begin{eqnarray} H &=& H_0 ~+~ V, \nonumber \\
H_0 &=& \sum_{n=1}^{N/2} ~[ J_1^A~\vec{ \mathbf{s}}_{2n-1} \cdot 
\vec{ \mathbf{s}}_{2n} ~-~ g\mu_B h~(\mathbf{s}_{2n}^z + 
\mathbf{s}_{2n-1}^z)], \nonumber \\
V &=& -\sum_{n=1}^{N/2}~ 
[J_2^F(\vec{\mathbf{s}}_{2n-1} \cdot \vec{ \mathbf{s}}_{2n+1}+
\vec{ \mathbf{s}}_{2n-2} \cdot \vec{ \mathbf{s}}_{2n}) \nonumber \\
&&~~~~~~~~~~~~~ +\vec{\mathbf{s}}_{2n-2}\cdot \vec{ \mathbf{s}}_{2n-1}]
\label{ham2} \end{eqnarray}
where $V$ will be treated perturbatively.

We first consider the unperturbed Hamiltonian $H_0$. The eigenstates of this
consist of decoupled pairs of spins at sites $(2n-1,2n)$; we will denote such 
a pair by the label $n$. $H_0$ commutes with
the total spin and the $z$ component of the total spin for each pair. For very
large values of $h$, the ground state of the system is unique and is given
by the fully polarized state with $s^z = s$ for all the spins;
the state of each pair of spins $n$ will then be given by
$|S_{tot},S_{tot}^z>$ = $|2s,2s>$ which in configuration basis
is $|s_{2n-1}^z, s_{2n}^z>$ = $|s,s>$. As $h$ is decreased, at a field
strength $h_0$, the ground state above will become degenerate with the state
$|2s-1,2s-1>$ which in configuration basis is given by $\frac{1}{\sqrt{2}}
[|s,s-1> - |s-1,s>]$. Using the expression for $H_0$, we find that 
the energy difference between these two states is $2s J_1^A - g\mu_B h$;
hence they become degenerate at
\begin{eqnarray} g \mu_B h_0 ~=~ 2s J_1^A. \label{h0} \end{eqnarray}
At this value of the field, one can check that the two-spin states with
$S_{tot} \le 2s -2$ are higher in energy by an amount proportional to the
magnetic field. We will therefore work only within the subspace of the two
states $(2s,2s)$ and $(2s-1,2s-1)$, and we will denote these as having
pseudo-spin $\tau_n^z = +1$ and $-1$, respectively.

We now effectively have a spin-1/2 chain. If $h$ differs slightly from the
value given in Eq.~\ref{h0}, we see that the effective Hamiltonian for the
spin-1/2 chain receives the contribution
\begin{eqnarray} \sum_n ~(sJ_1^A -\frac{g\mu_B h}{2}) ~\tau_n^z \end{eqnarray}
from $H_0$, ignoring a constant independent of $\tau_n^z$. We now have 
to add to this the contributions from first order in the perturbation $V$,
namely, the matrix elements $<\tau_{n}^z,\tau_{n+1}^z |V|\tau_{n}^{'z},
\tau_{n+1}^{'z}>$ corresponding to the various possibilities $\tau_n^z,
\tau_n^{'z},\tau_{n+1}^z,\tau_{n+1}^{'z} = \pm 1$. Adding all the terms and
ignoring some constants, we obtain the effective Hamiltonian
\begin{eqnarray} H_{eff} &=& \sum_n ~[ \frac{s}{2} (1- 2J_2^F) 
(\tau_n^+ \tau_{n+1}^- + \tau_n^-\tau_{n+1}^+) \nonumber \\
&& - \frac{1}{16} (1+2J_2^F) \tau_n^z \tau_{n+1}^z \label{heff}\\ 
&&+ (sJ_1^A - \frac{g\mu_B h}{2} - \frac{1}{8} (4s-1) (1+2J_2^F)) 
\tau_n^z]. \nonumber \end{eqnarray}
We thus obtain the Hamiltonian of a spin-1/2 $XXZ$ chain in a transverse
magnetic field. This has been studied extensively; see, for example, Refs.
\cite{johnson,alcaraz,cabra}. Note that our Hamiltonian is ferromagnetic 
since the coefficient of the $zz$ interaction is negative. The coefficient
of the $\tau_n^+ \tau_{n+1}^- + \tau_n^- \tau_{n+1}^+$ term can be made
negative, if it is not already so, by performing the unitary transformation
$\tau_n^{\pm} \to (-1)^n \tau_n^{\pm}$ and $\tau_n^z \to \tau_n^z$.

To simplify the notation for a while, let us write Eq.~\ref{heff} in the form
\begin{eqnarray} H_{eff} &=& \sum_n ~[~ - J ~(\tau_n^+ \tau_{n+1}^- + 
\tau_n^- \tau_{n+1}^+ ) \nonumber \\
& & ~~~~~~~- \Delta ~\tau_n^z \tau_{n+1}^z ~-~ \mu ~\tau_n^z], \label{heff2} 
\end{eqnarray}
where we assume that $\Delta > 0$.
The phase diagram of the Hamiltonian in Eq.~\ref{heff2} is known to be as
follows~\cite{johnson,alcaraz,cabra}. If $|J| < 2 \Delta$, the ground state is
given by all $\tau_n^z = +1$ if $\mu > 0$ and all $\tau_n^z = -1$ if $\mu < 0$.
Hence the ground state abruptly changes when $\mu$ crosses zero.
If $|J| > 2 \Delta$, the ground state changes from all $\tau_n^z = +1$ to all
$\tau_n^z = -1$ over a finite range of values of $\mu$. This range can be
found by the condition that the minimum energy $E_k$ of a spin wave (namely, 
a state with one $\tau_n^z = -1$ with amplitude $e^{ikn}$ while all the other 
$\tau_n^z = +1$, or one $\tau_n^z = +1$ while all the other $\tau_n^z = -1$) 
becomes equal to zero~\cite{tandon}. We find that the value of $\mu$
above which the ground state has all $\tau_n^z = +1$ is given by $\mu_+ = 
|J| - 2 \Delta$, while the value of $\mu$ below which the ground state has
all $\tau_n^z = -1$ is given by $\mu_- = - |J| + 2 \Delta$. Thus the average
value of $\tau_n^z$ per site will change gradually from $+1$ to $-1$ over a
range of values of $\mu$ given by $\mu_+ - \mu_- = 2|J| - 4 \Delta$.

We can now use the above results and map Eq.~\ref{heff2} back to
Eq.~\ref{heff} to obtain the phase diagram of our spin-$s$ chain in
a magnetic field. If
\begin{eqnarray} \frac{4s-1}{4s+1} ~<~ 2 J_2^F ~<~\frac{4s+1}{4s-1}, \label{j2}
\end{eqnarray}
we see that the ground state will abruptly change from all $\tau_n^z = +1$
to all $\tau_n^z = -1$ when $h$ crosses a value given by
\begin{eqnarray} g\mu_B h_c ~=~ 2sJ_1^A ~-~ \frac{1}{4} (4s-1) (1 + 2J_2^F). 
\end{eqnarray}
Returning to the original spin language, we see that the $S_{tot}^z$ changes
abruptly from $Ns$ for $h> h_c$ to $N(s-1/2)$ for $h< h_c$. On the other
hand, when $J_2^F$ lies outside the range given by Eq.~\ref{j2},
then $S_{tot}^z$ will change gradually from $Ns$ to $N(s-1/2)$ as $h$
decreases from $h_+$ to $h_-$, where
\begin{eqnarray}
g\mu_Bh_+ &=& 2sJ_1^A ~-~ \frac{1}{4} (4s-1) (1 + 2J_2^F) \nonumber \\
& & +~ s |1 - 2J_2^F| ~-~ \frac{1}{4} (1 + 2J_2^F), \nonumber \\
g\mu_Bh_- &=& 2sJ_1^A ~-~ \frac{1}{4} (4s-1) (1 + 2J_2^F) \nonumber \\
& & -~ s |1 - 2J_2^F| ~+~ \frac{1}{4} (1 + 2J_2^F). \end{eqnarray}

A comparison between the theoretical and numerical values of $h_c$ for some
representative points is shown in Table \ref{tbl1}. We see that the
agreement between the theoretical and numerical values is better if $J_1^A$
is larger, since the perturbative derivation of the effective Hamiltonian
given above is more justified in that case.

\begin{table}[]
\begin{center}
\caption{Comparison between theoretical and numerical $h_c$. We have used
the value $g \mu_B$ = 1.34 K/Tesla to convert the values of $h$ in Tesla
to the values of the exchange constants in K.}
\vspace{0.6cm}

\begin{tabular}{|l|l|r|c|c|}
\hline
 & $J_1^A$ & $J_2^F$ & Theoretical $h_c$ & Numerical $h_c$ \\ \hline
\multirow{2}{*}{$s$ = 1/2 }& 1.2 & 0.8 & 0.409 & 0.425 \\
 & 1.4 & 0.5 & 0.667 & 0.675 \\ \hline
\multirow{2}{*}{$s$ = 1 }& 1.2 & 0.8 & 0.335 & 0.525 \\
 & 1.4 & 0.5 & 0.968 & 1.035 \\
\hline
\end{tabular}
\label{tbl1}
\end{center}
\end{table}

\section{Conclusion}
We have made a detailed study of a frustrated Heisenberg spin chain with
alternating ferromagnetic and antiferromagnetic nearest-neighbor exchanges 
and a next-nearest-neighbor ferromagnetic exchange. While some features of 
the spin-1/2 model had been studied earlier, we are the first to study the 
spin-1 model. We have shown that the ground state of this model can be found 
exactly for the spin-1/2 and spin-1 cases along a particular line in the 
parameter space, and we subsequently conjectured that the ground state is 
exactly solvable on that particular line for any site spin. The complete 
quantum phase diagram, found numerically, has a ferromagnetic and a 
non-magnetic phase as predicted by a classical analysis; in fact, the 
phase diagram for spin-1 is in much better agreement with the classical phase 
diagram than for spin-1/2. A study of the structure factor 
shows that the non-magnetic phase consists of three regions, called N\'eel, 
double-period N\'eel and spiral; however, these are separated from each other 
only by cross-over regions rather than true phase transitions.

The energy gap and entanglement entropy have been
studied to shed new light on the non-magnetic phase. It is found 
that while the non-magnetic phase for the spin-1/2 system is totally gapped, 
there is a gapless region in the otherwise gapped non-magnetic phase for the 
spin-1 system. In the presence of a magnetic field, the system exhibits 
macroscopic magnetization jumps for both the spin-1/2 and spin-1 cases. 
We have studied this phenomena using an effective Hamiltonian.

\begin{center}
{\large\bf Acknowledgements}
\end{center}
\noindent S. R. is thankful to DST, India for financial support through various
projects. D. S. thanks DST, India for support under SR/S2/JCB-44/2010.

\end{document}